\documentclass[conference]{IEEEtran}
\IEEEoverridecommandlockouts
\usepackage{amsmath}
\usepackage{amsfonts}
\usepackage{algorithmic}
\usepackage{algorithm}
\usepackage{array}
\usepackage[caption=true,font=normalsize,labelfont=rm,textfont=rm]{subfig}
\usepackage{textcomp}
\usepackage{url}
\usepackage{hyperref}
\usepackage{verbatim}
\usepackage{graphicx}
\usepackage{caption}
\usepackage{cite}
\usepackage[font=scriptsize,skip=5pt]{caption}
 \usepackage{balance}
 \def\BibTeX{{\rm B\kern-.05em{\sc i\kern-.025em b}\kern-.08em
    T\kern-.1667em\lower.7ex\hbox{E}\kern-.125emX}}
    
\usepackage{float}
\usepackage[utf8]{inputenc}
\usepackage{pgfplots}
\DeclareUnicodeCharacter{2212}{−}
\usepgfplotslibrary{groupplots,dateplot}
\usetikzlibrary{patterns, shapes.arrows}
\pgfplotsset{compat=newest}

\begin{document}

\title{Learning Short Codes for Fading Channels with No or Receiver-Only Channel State Information}
\author{
    \IEEEauthorblockN{Rishabh Sharad Pomaje\IEEEauthorrefmark{1} and Rajshekhar V Bhat\IEEEauthorrefmark{2}}
    \IEEEauthorblockA{Indian Institute of Technology Dharwad, Dharwad, Karnataka, India \\
    Email: 
    \{\IEEEauthorrefmark{1}210020036,   \IEEEauthorrefmark{2}rajshekhar.bhat\}@iitdh.ac.in}}
	\maketitle
 \pagenumbering{arabic} 
 
\begin{abstract}

In next-generation wireless networks, low latency often necessitates short-length codewords that either do not use channel state information (CSI) or rely solely on CSI at the receiver (CSIR). Gaussian codes that achieve capacity for AWGN channels may be unsuitable for \textit{no-CSI} and \textit{CSIR-only} cases. In this work, we design short-length codewords for these cases using an autoencoder architecture. From the designed codes, we observe the following: In the no-CSI case,  the learned codes are mutually orthogonal when the distribution of the real and imaginary parts of the fading random variable has support over the entire real line. However, when the support is limited to the non-negative real line, the codes are not mutually orthogonal. 
For the CSIR-only case, deep learning-based codes designed for AWGN channels perform worse in fading channels with optimal coherent detection compared to codes specifically designed for fading channels with CSIR, where the autoencoder jointly learns encoding, coherent combining, and decoding. In both no-CSI and CSIR-only cases, the codes perform at least as well as or better than classical codes of the same block length.

\end{abstract}

\section{Introduction}
Modern communication networks transmit information by varying electromagnetic wave characteristics emitted by antennas. These networks are based on electromagnetic theory, information theory, wireless propagation modeling, and antenna theory \cite{EIT}. With the transition to sixth generation (6G) networks, covering nearly all frequency bands and global environments, integrating these theories is crucial \cite{EIT,6gSurveyPaper}. However, doing so may not yield tractable models for wireless channels which are impaired by time-varying fading. Therefore, it is crucial to develop communication strategies that can operate effectively in fading channels.

% Modern communication networks transmit information by varying the characteristics of electromagnetic waves emitted by antennas. These networks are built on electromagnetic theory, information theory, wireless propagation modeling, and antenna theory which have historically evolved independently \cite{EIT}. As we advance to 6G wireless communication systems, operations will span nearly all frequency bands (from sub-6 GHz to optical) with global coverage across terrestrial, space, air, and sea environments, meeting the demands for higher data rates, low latency, and goal-oriented communication \cite{EIT,6gSurveyPaper}. In these scenarios, integrating these four theories is essential, but doing so may not yield well-defined models for wireless channels, whose fading coefficients vary across space and time. Therefore, it is crucial to develop codewords that can operate effectively in fading channels.

For communication over fading channels, techniques like interleaving and diversity using perfect channel state information (CSI) at the receiver (CSIR) and/or transmitter, are first employed to mitigate the effects of fading to make the channel resemble an additive white Gaussian noise (AWGN)  channel. Then, codewords designed for AWGN channels are adopted \cite{CodingforFadingChannelSurvey}. 
This is supported by studies such as \cite{FadingConverges2AWGNwithDiversity,SpreadResponseCode}, which suggest that Gaussian codes, capacity-achieving for AWGN channels, also perform well on fading channels asymptotically with infinite interleaving depth and receiver diversity. However, achieving infinite interleaving depth and diversity is impractical due to computational, hardware, and latency constraints. Thus, finite and short blocklength codewords are needed. Designing good short-length codewords is challenging even in AWGN channels, and their performance degradation can be particularly severe in fading environments.

As mentioned in \cite{HowPerfect}, with asymptotic diversity, capacity is achieved in fading channels using Gaussian codebooks and scaled nearest-neighbor decoding. However, capacity is highly sensitive to channel estimation errors, which are particularly challenging to control in dynamic 6G scenarios. Given the overhead associated with channel estimation, communicating without CSI (or with only CSIR) may be more practical, especially in 6G where the number of antennas is large and fading varies rapidly \cite{yang2019capacityfadingchannelschannel}.  Various works have explored signal design and capacity derivation for fading channels without CSI. \cite{Unitary} introduces unitary space-time modulation for multiple-antenna links without CSI in Rayleigh block-fading channels, where fading coefficients remain constant over multiple symbol periods. For such channels, \cite{GrassmannManifold} derives a capacity expression without CSI, which can be interpreted geometrically as sphere packing in the Grassmann manifold. Moreover, \cite{Discrete-Distribution,Goldsmith-NoCSICapacity} demonstrate that the capacity-achieving distribution for Rayleigh block-fading fading channels without CSI is discrete.
Despite these advancements, the development of efficient short-length codewords for Rayleigh and other fading channels under No-CSI and CSIR-Only cases remains a challenging and under-explored area. 

Deep learning can be used to design \textit{good} codewords for finite blocklengths in AWGN and fading channels.  Previous works \cite{IntroDeepPhysical,erpek2020deepWireless} show that deep learning-based codewords perform as well as or better than traditional codes in AWGN channels. However, there has been limited exploration of codeword design using deep learning for fading channels, possibly due to the assumption that AWGN-optimized codes would also perform well in fading scenarios, even for finite block lengths.  In this work, we assume a fading channel with fixed statistics. Towards addressing the above gaps, our contributions are:
%Motivated by the observation that codes designed for AWGN channels may not be suitable for fading environments without CSI or with only CSIR, our contributions are:

% Deep learning can be used to design effective codewords for finite block lengths in AWGN and fading channels. Previous works \cite{IntroDeepPhysical,erpek2020deepWireless} show that deep learning-based codewords perform as well as or better than traditional codes in AWGN channels. However, there is limited exploration of deep learning for fading channels, possibly due to the assumption that AWGN-optimized codes would also perform well in fading scenarios. This work assumes a fading channel with fixed statistics and addresses the limitations of AWGN-optimized codes in fading environments without CSI or with only CSIR. Our contributions are:

\emph{We design short-length codes specifically for fading channels under no-CSI and CSIR-only cases. In the no-CSI case, we observe that the learned codes are mutually orthogonal when the distribution of the real and imaginary parts of the fading random variable has support over the entire real line, \(\mathbb{R}\). However, when the support is limited to the non-negative real line, \(\mathbb{R}_+\), the codes are not mutually orthogonal. For the CSIR-only case, deep learning-based codes designed for AWGN channels perform worse when applied to fading channels with optimal coherent detection compared to codes directly designed for fading channels with CSIR. In the latter case, the deep learning-based autoencoder jointly learns encoding, coherent combining, and decoding. In both no-CSI and CSIR-only scenarios, these codes perform at least as well as or better than classical codes of the same block length.}

%However, fading channels differ from AWGN channels due to their non-memoryless nature and variable signal-to-noise ratio (SNR). .    True memorylessness requires infinite interleaving depth.  
% Moreover, knowledge of channel state information (CSI) is crucial for designing and evaluating codewords. 

%Consequently, codes designed for AWGN channels may perform poorly without CSI, regardless of codeword length.    
% May be useful: https://ieeexplore.ieee.org/stamp/stamp.jsp?tp=&arnumber=485720 
%as with finite blocklengths, we cannot convert the fading channel into AWGN-like channel, by reducing the probability of channel being in deep fade. 

% Beyond the scope of this paper, it may be advantageous to design codes specifically for fading channels due to additional points in our initial motivation, which include: (i) the fact that codes optimized for AWGN channels may not be suitable for fading channels due to their memoryless nature, where past states influence the current state, and (ii) the varying propagation characteristics of fading channels over time, which necessitate adaptive coding strategies. These aspects warrant further investigation, as discussed in \cite{CodingforFadingChannelSurvey}, which we consider for future work.

    \subsection{Literature Survey}
    In \cite{IntroDeepPhysical}, an autoencoder with an encoder, noise layer, and decoder is used. The encoder maps message indices to norm-constrained codewords,   the noise layer simulates an AWGN channel and the decoder recovers the original message from the noisy codeword. This approach performs comparably to classical codes. Further research has applied similar techniques to degraded broadcast channels \cite{stauffer2019Broadcast} and multiple-input multiple-output (MIMO) channels \cite{oshea2017deepMIMO}.

Deep learning has also been integrated with classical communication systems to enhance receivers \cite{neumann2018MMSE, farshad2018neuralDetectors} and detect sequential codes like convolutional and turbo codes \cite{kim2020PhysicalDeep}. Additionally, it has been applied to channel estimation in frequency-selective fading channels \cite{Ye2018deepOFDM}. In designing communication systems over fading channels, scenarios typically considered include (i) perfect CSI at the receiver (CSIR) but not at the transmitter (CSIT), and (ii) perfect CSIT and CSIR.  \cite{oshea2017deepMIMO} addresses communication over fading channels with perfect CSIR and CSIT, including scenarios with quantized CSI.

Training an autoencoder end-to-end is challenging without instantaneous channel transfer function knowledge, as the channel must be modeled in intermediate layers, and backpropagation requires functional forms for all layers. Methods to address the above challenge include Simultaneous Perturbation Stochastic Optimization (SPSA) \cite{Raj2018BkpropAir} and Generative Adversarial Networks (GANs) \cite{Ye2020WirelessGAN}, where a GAN is trained with encoded signals and pilot data to serve as a surrogate channel for training the transmitter and receiver DNNs.

Following \cite{IntroDeepPhysical}, we use an autoencoder to learn short codes for fading channels. However, to the best of our knowledge, none of the above works address the no-CSI and CSIR-only cases in detail, which we focus on.

    \section{System Model and Methodology}
    
    We consider a point-to-point communication system, where the transmitter communicates information over a wireless fading channel. Specifically, let \( x[l] \) denote the complex symbol transmitted in the \( l \)-th channel use. The receiver observes 
    \begin{align}
        y[l] &= h[l]x[l] + w[l], \label{eq-channel}
    \end{align}
    where, \( w[l] \) represents AWGN with mean $0$ and variance \( N_0 \), and \( h[l] \) is the complex channel gain, with variance unity.  
    We assume that \( w[1], w[2], \ldots \) are independent and identically distributed (i.i.d.) circularly symmetric complex normal (CSCN)    random variables, i.e., $w[l] \overset{\text{i.i.d.}}{\sim} \mathcal{CSCN}(0, N_0)$.

    We are interested in mapping a message \( m \in \mathcal{M} \), where \( \mathcal{M} \) represents the set of all possible messages with cardinality, $|\mathcal{M}|=M$, to a codeword \( \mathbf{c} = [c_1, \ldots, c_n] \), where \( c_i \in \mathbb{R} \) and \( n \) is the length of the codeword. Each codeword is expected to satisfy an energy constraint: $\|\mathbf{c}\|_2^2 \leq n$.
    Let \( \mathcal{C} = \{\mathbf{c} \mid \mathbf{c} \text{ is a codeword for } m \in \mathcal{M}\} \) be the codebook. The mapping function \( f_{\Theta}: \mathcal{M} \rightarrow \mathcal{C} \) maps each message \( m \in \mathcal{M} \) to a unique codeword. A codeword \( \mathbf{c} \) is transmitted over a fading channel across consecutive time slots, such that \( x[1] = c_1 \), \( x[2] = c_2 \), \ldots, \( x[n] = c_n \), after which the next codeword is transmitted. At the receiver, all symbols corresponding to a codeword are collected, resulting in \( \mathbf{y} = [y[1], \ldots, y[n]] \). The goal is to estimate the transmitted message \( m \) using another function, \( g_{\Phi}: \mathbb{R}^n \rightarrow \mathcal{M} \). Thus, \( \hat{m} = g_{\Phi}(\mathbf{y}) \) denotes the estimate of the transmitted message.
    Here, \( \Theta \) and \( \Phi \) are the parameters of the encoding and decoding functions, which we intend to learn. The functions are optimized to minimize the block error rate (BLER), defined as 
$        \epsilon(M,n) =  (1/M)\sum_{m \in \mathcal{M}} \Pr\left(g_{\Phi}(\mathbf{y}) \neq m \mid \mathbf{x} = f_{\Theta}(m)\right),
 $ where the relationship between \( \mathbf{y} \) and \( \mathbf{x} \) is given by \eqref{eq-channel}. 
  We  empirically compute the BLER as follows:
    \begin{align}\label{eq-epsilon}
         \hat{\epsilon}(M,n) &= \frac{1}{N} \sum_{i=1}^{N} I(m^{(i)}, \hat{m}^{(i)}),
    \end{align}
    where \( I \) is the indicator function defined as \( I(m, \hat{m}) = 0 \) if \( m = \hat{m} \), and \( I(m, \hat{m}) = 1 \) otherwise, and \( N \) is the total number of messages sampled uniformly at random from \( \mathcal{M} \).  We denote the BLER  as \( \epsilon_{\mathrm{No CSI}}(M,n) \) and  $\epsilon_{\mathrm{CSIROnly}}(M,n)$ for no-CSI and CSIR-Only cases, respectively.

%    As mentioned, in this work, we consider various degrees of knowledge about channel state information (CSI), specifically the knowledge of instantaneous \( h[l] \) at the transmitter and the receiver. When there is no CSI at either the transmitter or the receiver, . When there is CSI at the receiver (CSIR) but no CSI at the transmitter (CSIT), we denote it as \( \epsilon_{\mathrm{Only CSIR}}(M,n) \). Finally, when both CSIT and CSIR are available, we write it as \( \epsilon_{\mathrm{Full CSI}}(M,n) \).    

We denote the energy per coded bit as \( E_b \) and define the signal-to-noise ratio (SNR) as \( \text{SNR}_{\text{linear}} = E_b / N_0 \) and \( \text{SNR}_{\text{dB}} = 10 \log_{10} \text{SNR}_{\text{linear}} \). With \( E_b = 1 \), we adjust the noise variance for different SNRs as follows: For the uncoded case, \( N_0 = 1 / (2 \times \text{SNR}_{\text{linear}}) \). For the coded case, \( N_0 = 1 / (2 \times R \times \text{SNR}_{\text{linear}}) \), where \( R = \log_2 M / n \). 
%The noise variance applies to individual i.i.d. real or imaginary components.

We use an autoencoder to learn the functions \( f_{\Theta} \) and \( g_{\Phi} \) to minimize \eqref{eq-epsilon}. The architecture, hyperparameters, models, and source code are available\footnote{Access the source code \href{https://github.com/RishP11/Learning-Short-Codes-for-Fading-Channels-with-No-or-Receiver-Only-Channel-State-Information}{here on \texttt{GitHub}}.}. The autoencoder's input and expected output are one-hot vectors of dimension \( M \times 1 \). Both functions \( f_{\Theta} \) and \( g_{\Phi} \) are feedforward neural networks. The encoder's final layer normalizes the output vector to unit norm, which is then scaled by the fading coefficient and perturbed with Gaussian noise. The decoder outputs a probability mass function with \( M \) nodes using a softmax activation function. The loss is the cross-entropy between the one-hot input vector and the decoder’s output probabilities.

\section{Results and Discussion}
We now present the results. The energy per codeword is set to $n$ by design. We adjust the noise variance accordingly to obtain the probability of error versus SNR in the figures.

\subsection{No CSI Case}
We consider both Rayleigh  and non-Rayleigh fading cases. 
\subsubsection{Rayleigh Fading}
We now consider the Rayleigh fading case, where  $h[l] = h_r[l]+jh_i[l] \overset{\text{i.i.d.}}{\sim} \mathcal{CSCN}(0, 1)$. We consider different values of \( M\) with varying \( n \) values. 
The autoencoder's performance is compared to uncoded orthogonal signaling, mentioned in \cite{Tse_Viswanath_2005}, where the transmitter sends \( [c_1, c_2] \) equal to \( [1, 0] \) and \( [0, 1] \) for bits $0$ and $1$, respectively.  At the decoder, maximum likelihood decoding (MLD) determines bit $0$ if \( |y_1| > |y_2| \), where \( [y_1, y_2] \) is the received vector from these consecutive symbols.

    \begin{center}
        \begin{table}[t]
            \caption{Codebook learned for $M=2$ case for different values of $n$.   Each element is rounded to two decimal places. } 
            \label{tab-CodebookM2}
            \begin{tabular}{|c|p{7.6cm}|}
                \hline
                $n$ &  Codebook \\
                \hline
                $2$ & $[1.41, 0.0], [0.0, -1.41]$ \\
                \hline
                $3$ & $[0.0, 0.0, -1.73],[1.73, 0.0, 0.0]$ \\
                \hline
                $4$ & $[0.0, 0.0, 0.0, 1.99], [1.16, -1.17, -1.14, 0.0]$ \\
                \hline
                $5$ & $[0.0, 0.0, 1.59, 0.0, 1.57], [1.59, 0.0, 0.0, -1.57, 0.0]$ \\
                \hline
            \end{tabular}
        \end{table}
    \end{center} 
    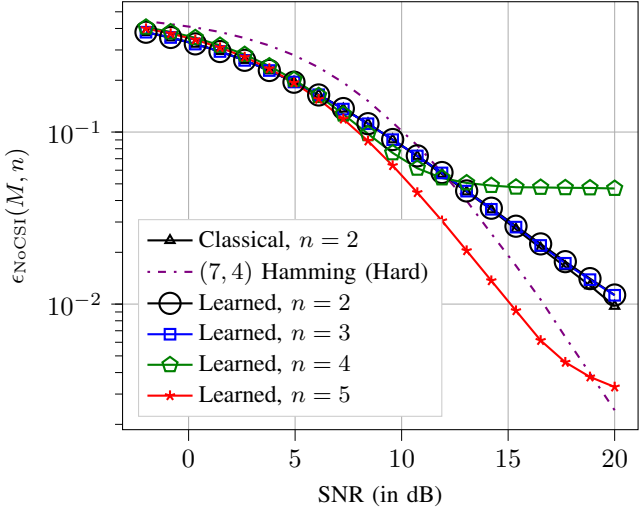
\begin{figure}[t]
        \centering
        % This file was created with tikzplotlib v0.10.1.
\begin{tikzpicture}

\definecolor{darkgray176}{RGB}{176,176,176}
\definecolor{green}{RGB}{0,128,0}
\definecolor{lightgray204}{RGB}{204,204,204}
\definecolor{purple}{RGB}{128,0,128}

\begin{axis}[
legend cell align={left},
legend style={
  fill opacity=0.8,
  draw opacity=1,
  text opacity=1,
  at={(0.03,0.03)},
  anchor=south west,
  draw=lightgray204
},
log basis y={10},
tick align=outside,
tick pos=left,
x grid style={darkgray176},
xlabel={\small SNR (in dB)},
xmajorgrids,
xmin=-3.1, xmax=21.1,
xtick style={color=black},
y grid style={darkgray176},
ylabel={\small \(\displaystyle \epsilon_{\mathrm{No CSI}}(M,n)\)},
ymajorgrids,
ymin=0.00186301261741955, ymax=0.572208873967023,
ymode=log,
ytick style={color=black}
]
\addplot [thick, black, mark=triangle, mark size=2, mark options={solid}]
table {%
-2 0.379885
-0.842105263157895 0.353048
0.315789473684211 0.32461
1.47368421052632 0.29466
2.63157894736842 0.261365
3.78947368421053 0.22714
4.94736842105263 0.195263
6.10526315789474 0.164991
7.26315789473684 0.136231
8.42105263157895 0.111861
9.57894736842105 0.090276
10.7368421052632 0.072357
11.8947368421053 0.056944
13.0526315789474 0.044984
14.2105263157895 0.035383
15.3684210526316 0.027273
16.5263157894737 0.021171
17.6842105263158 0.016492
18.8421052631579 0.012801
20 0.00973
};
\addlegendentry{\small Classical, $n=2$}
\addplot [thick, purple, dash pattern=on 1pt off 3pt on 3pt off 3pt]
table {%
-2 0.441056
-0.842105263157895 0.423869
0.315789473684211 0.405249
1.47368421052632 0.381611
2.63157894736842 0.351994
3.78947368421053 0.318561
4.94736842105263 0.279774
6.10526315789474 0.237635
7.26315789473684 0.193919
8.42105263157895 0.151703
9.57894736842105 0.113621
10.7368421052632 0.08356
11.8947368421053 0.058248
13.0526315789474 0.039873
14.2105263157895 0.025686
15.3684210526316 0.016703
16.5263157894737 0.010626
17.6842105263158 0.006438
18.8421052631579 0.003953
20 0.002417
};
\addlegendentry{\small $(7, 4)$ Hamming (Hard)}
\addplot [thick, black, mark=o, mark size=4, mark options={solid}]
table {%
-2 0.380393
-0.842105263157895 0.354656
0.315789473684211 0.325443
1.47368421052632 0.293783
2.63157894736842 0.261278
3.78947368421053 0.227956
4.94736842105263 0.194866
6.10526315789474 0.164265
7.26315789473684 0.136946
8.42105263157895 0.111854
9.57894736842105 0.090471
10.7368421052632 0.072734
11.8947368421053 0.058164
13.0526315789474 0.045473
14.2105263157895 0.036089
15.3684210526316 0.028363
16.5263157894737 0.022382
17.6842105263158 0.01765
18.8421052631579 0.014066
20 0.011295
};
\addlegendentry{\small Learned, $n=2$}
\addplot [thick, blue, mark=square, mark size=2, mark options={solid,fill opacity=0}]
table {%
-2 0.38057
-0.842105263157895 0.354521
0.315789473684211 0.324658
1.47368421052632 0.293878
2.63157894736842 0.26103
3.78947368421053 0.227845
4.94736842105263 0.195436
6.10526315789474 0.164789
7.26315789473684 0.135989
8.42105263157895 0.11176
9.57894736842105 0.090314
10.7368421052632 0.072659
11.8947368421053 0.057859
13.0526315789474 0.04535
14.2105263157895 0.035566
15.3684210526316 0.028014
16.5263157894737 0.021924
17.6842105263158 0.01727
18.8421052631579 0.013873
20 0.011252
};
\addlegendentry{\small Learned, $n=3$}
\addplot [thick, green, mark=pentagon, mark size=3, mark options={solid}]
table {%
-2 0.406482
-0.842105263157895 0.381905
0.315789473684211 0.353347
1.47368421052632 0.318946
2.63157894736842 0.282176
3.78947368421053 0.242002
4.94736842105263 0.201669
6.10526315789474 0.162784
7.26315789473684 0.126985
8.42105263157895 0.09765
9.57894736842105 0.07565
10.7368421052632 0.061568
11.8947368421053 0.054065
13.0526315789474 0.050596
14.2105263157895 0.048862
15.3684210526316 0.047785
16.5263157894737 0.047657
17.6842105263158 0.04743
18.8421052631579 0.047321
20 0.046987
};
\addlegendentry{\small Learned, $n=4$}
\addplot [thick, red, mark=star, mark size=2, mark options={solid,fill opacity=0}]
table {%
-2 0.399246
-0.842105263157895 0.373794
0.315789473684211 0.343976
1.47368421052632 0.309839
2.63157894736842 0.272917
3.78947368421053 0.233577
4.94736842105263 0.193395
6.10526315789474 0.155195
7.26315789473684 0.119054
8.42105263157895 0.088778
9.57894736842105 0.063855
10.7368421052632 0.044493
11.8947368421053 0.030485
13.0526315789474 0.020446
14.2105263157895 0.013678
15.3684210526316 0.009157
16.5263157894737 0.006148
17.6842105263158 0.004591
18.8421052631579 0.003774
20 0.003298
};
\addlegendentry{\small Learned, $n=5$}
\end{axis}

\end{tikzpicture}
        \caption{BLER for \(M=2\) with varying \(n\). The training SNR is \(7\) dB.}
        \label{fig-multi_n_noCsi_M2}
    \end{figure}

The codewords obtained using the autoencoder and their performance are shown in Table 
 \ref{tab-CodebookM2} and Fig.\ref{fig-multi_n_noCsi_M2} for $M=2$, and in Table~\ref{tab-CodebookM4} and Fig.~\ref{fig-multi_n_noCsi_M4} for $M=4$, respectively. As observed in the tables, the energy per codeword is designed to equal $n$, and, importantly, the codewords are mutually orthogonal for each $n$. In these figures, for the classical $n=2$ case, we use the classical orthogonal signaling mentioned above. For the $(7,4)$ Hamming code, we first encode the bits and then apply orthogonal signaling for transmission, and at the receiver, we first detect the bits using the aforementioned threshold algorithm, which gives hard detection of bits and then proceed with syndrome decoding.

% The motivation for exploring different \( M \) and \( n \) stems from the observation that for \( M = 2 \) and \( n = 2 \), the autoencoder's performance matched that of classical orthogonal signaling, with the encoder learning the same orthogonal mapping, where its \( 2 \times 1 \) output represented two symbols over two durations. The decoder, like MLD, compared absolute values. Extending orthogonal signaling to larger \( n \) and \( M \) is non-trivial, but we found that the learned codewords remained orthogonal, with performance improving as \( n \) increased. For \( n = 1 \) and \( M = 2 \), orthogonal codewords with equal energy are not possible, limiting the probability of error to 0.5 unless the equal energy constraint is relaxed. In general, no signaling scheme works for \( n < M \) in the no-CSI case, as we cannot generate more than \( n \) mutually orthogonal vectors in \( n \)-dimensional space.

The motivation for exploring different \( M \) and \( n \) stems from the observation that for \( M = 2 \) and \( n = 2 \), the autoencoder's performance matched that of classical orthogonal signaling, with the encoder learning the same orthogonal mapping. 
%, where its \( 2 \times 1 \) output represented two symbols over two durations. 
The decoder, like MLD, compared absolute values. Extending orthogonal signaling to larger \( n \) and \( M \) is non-trivial, but we found that the learned codewords remained orthogonal, with performance improving as \( n \) increased.  For \( n = 1 \) and \( M = 2 \), it is not possible to find orthogonal codewords with equal energy. In general, no signaling scheme works for \( n < M \), as we cannot generate more than \( n \) mutually orthogonal vectors in \( n \)-dimensional space.

% The motivation for obtaining the results for different $M$ and $n$ is the following: We observed that for $M= 2$ and $n=2$, the autoencoder's performance matched that of the classical orthogonal signaling scheme. The encoder learned the same orthogonal mapping, with its \( 2 \times 1 \) output interpreted as two symbols over two durations. It is observed that the decoder compared absolute values, similar to the MLD. Extending orthogonal signaling to larger \( n \) values for different $M$ values is non-trivial, before we observe that the learned codewords are mutually orthogonal.  For this, we train the autoencoder for various \( n \) and observe that the learned codewords are orthogonal and also observe  improved performance with increasing \( n \). For \( n=1 \) and $M=2$ case, since each codeword must have the same energy by design and we cannot get two such codewords which are orthogonal while having the same energy, and hence, no signaling and decoding schemes will offer probability of error less than $0.5$ (if the requirement of equal energy in our design is relaxed, a signaling scheme in which we will transmit symbol $0$ for bit $0$ and some non-zero symbol for bit $1$ or shifted version of them, would yield a probability of error $<0.5$.) In general, since in an $n$-dimensional space we cannot get more than $n$ mutually orthogonal vectors, no signaling scheme works for $n<M$ in the no CSI case. 

    \begin{center}
        \begin{table}[t]
            \centering
            \caption{Codebook learned for $M=4$ case for different values of $n$.   Each element is rounded to two decimal places. } 
            \label{tab-CodebookM4}
            \begin{tabular}{|c|p{8cm}|}
                \hline
                $n$ & Codebook \\
                \hline
                $4$ & \begin{tabular}{@{}l@{}}
                    $[0.00, 2.0, 0.00, 0.00],[0.00, 0.00, 0.00, -2.0],$ \\
                    $[2.0, 0.01, 0.01, 0.01], [0.00, 0.00, -2.0, 0.01]$
                \end{tabular} \\
                \hline 
                $5$ & 
                \begin{tabular}{@{}l@{}}
                    $[-2.24, 0.0, 0.0, 0.01, 0.0], [0.0, 2.24, 0.0, 0.0, 0.0],$ \\
                    $[0.02, 0.0, 0.0, 0.01, 2.24], [0.0, 0.0, 0.0, 2.24, 0.0]$
                \end{tabular} \\
                \hline
                $6$ & 
                \begin{tabular}{@{}l@{}}
                    $[0.00, 0.00, 0.00, 0.00, 2.4, 0.00], [0.01, 0.00, 2.4, 0.00, 0.00, 0.00],$ \\
                    $[2.4, -0.01, 0.00, 0.00, 0.01, 0.00], [0.00, -2.4, 0.0, 0.0, 0.01, 0.00]$
                \end{tabular} \\
                % \hline
                % $7$ & 
                % \begin{tabular}{@{}l@{}}
                %     $[-2.6, 0.00, 0.01, 0.00, 0.00, 0.00, 0.00],$ \\
                %     $[0.00, 0.00, -2.6, 0.00, 0.00, 0.00, 0.00],$ \\
                %     $[0.00, 0.00, 0.00, 0.00, 0.00, -2.6, -0.01],$ \\
                %     $[0.00, 0.01, 0.01, 1.9, 0.00, 0.01, 1.9]$
                % \end{tabular} \\
                % \hline
                % $8$ & 
                % \begin{tabular}{@{}l@{}}
                %     $[-0.02, -1.6, -1.6, 0.00, 0.00, -0.02, 1.6, 0.00],$ \\
                %     $[-2.0, 0.00, 0.01, -2.0, -0.01, 0.01, 0.01, 0.00],$ \\
                %     $[-0.02, -0.01, 0.02, -0.00, 0.00, -2.8, 0.01, 0.00],$\\
                %     $[-0.01, 0.00, 0.00, -0.01, 2.8, 0.00, 0.02, 0.01]$
                % \end{tabular} \\
                \hline
            \end{tabular}
        \end{table}               
    \end{center}
    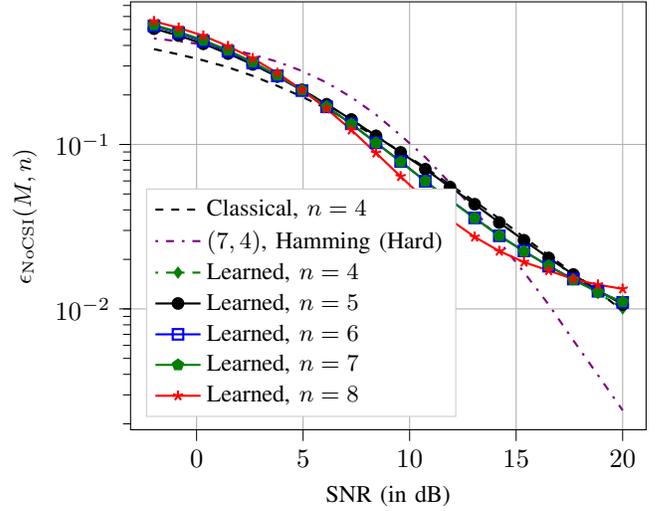
\begin{figure}[t]
        \centering
        % This file was created with tikzplotlib v0.10.1.
\begin{tikzpicture}

\definecolor{darkgray176}{RGB}{176,176,176}
\definecolor{green}{RGB}{0,128,0}
\definecolor{lightgray204}{RGB}{204,204,204}
\definecolor{purple}{RGB}{128,0,128}

\begin{axis}[
legend cell align={left},
legend style={
  fill opacity=0.8,
  draw opacity=1,
  text opacity=1,
  at={(0.03,0.03)},
  anchor=south west,
  draw=lightgray204
},
log basis y={10},
tick align=outside,
tick pos=left,
x grid style={darkgray176},
xlabel={\small SNR (in dB)},
xmajorgrids,
xmin=-3.1, xmax=21.1,
xtick style={color=black},
y grid style={darkgray176},
ylabel={\small \(\displaystyle \epsilon_{\mathrm{No CSI}}(M,n)\)},
ymajorgrids,
ymin=0.00184071607674548, ymax=0.736823374410902,
ymode=log,
ytick style={color=black}
]
\addplot [thick, black, dashed]
table {%
-2 0.379885
-0.842105263157895 0.353048
0.315789473684211 0.32461
1.47368421052632 0.29466
2.63157894736842 0.261365
3.78947368421053 0.22714
4.94736842105263 0.195263
6.10526315789474 0.164991
7.26315789473684 0.136231
8.42105263157895 0.111861
9.57894736842105 0.090276
10.7368421052632 0.072357
11.8947368421053 0.056944
13.0526315789474 0.044984
14.2105263157895 0.035383
15.3684210526316 0.027273
16.5263157894737 0.021171
17.6842105263158 0.016492
18.8421052631579 0.012801
20 0.00973
};
\addlegendentry{\small Classical, $n=4$}
\addplot [thick, purple, dash pattern=on 1pt off 3pt on 3pt off 3pt]
table {%
-2 0.441056
-0.842105263157895 0.423869
0.315789473684211 0.405249
1.47368421052632 0.381611
2.63157894736842 0.351994
3.78947368421053 0.318561
4.94736842105263 0.279774
6.10526315789474 0.237635
7.26315789473684 0.193919
8.42105263157895 0.151703
9.57894736842105 0.113621
10.7368421052632 0.08356
11.8947368421053 0.058248
13.0526315789474 0.039873
14.2105263157895 0.025686
15.3684210526316 0.016703
16.5263157894737 0.010626
17.6842105263158 0.006438
18.8421052631579 0.003953
20 0.002417
};
\addlegendentry{\small $(7, 4)$, Hamming (Hard)}
\addplot [thick, green, dash pattern=on 1pt off 3pt on 3pt off 3pt, mark=diamond*, mark size=1.75, mark options={solid}]
table {%
-2 0.506617
-0.842105263157895 0.459159
0.315789473684211 0.40824
1.47368421052632 0.356912
2.63157894736842 0.307726
3.78947368421053 0.258289
4.94736842105263 0.214796
6.10526315789474 0.176179
7.26315789473684 0.14161
8.42105263157895 0.113334
9.57894736842105 0.090036
10.7368421052632 0.071042
11.8947368421053 0.055585
13.0526315789474 0.043314
14.2105263157895 0.033639
15.3684210526316 0.026158
16.5263157894737 0.020498
17.6842105263158 0.015943
18.8421052631579 0.012495
20 0.010075
};
\addlegendentry{\small Learned, $n=4$}
\addplot [thick, black, mark=*, mark size=2, mark options={solid}]
table {%
-2 0.506175
-0.842105263157895 0.459575
0.315789473684211 0.409474
1.47368421052632 0.357076
2.63157894736842 0.306748
3.78947368421053 0.258797
4.94736842105263 0.214852
6.10526315789474 0.175402
7.26315789473684 0.141764
8.42105263157895 0.112681
9.57894736842105 0.089428
10.7368421052632 0.070697
11.8947368421053 0.055064
13.0526315789474 0.043276
14.2105263157895 0.03349
15.3684210526316 0.026126
16.5263157894737 0.0204
17.6842105263158 0.016185
18.8421052631579 0.012949
20 0.010635
};
\addlegendentry{\small Learned, $n=5$}
\addplot [thick, blue, mark=square, mark size=2, mark options={solid,fill opacity=0}]
table {%
-2 0.525873
-0.842105263157895 0.47726
0.315789473684211 0.424285
1.47368421052632 0.369047
2.63157894736842 0.313283
3.78947368421053 0.261211
4.94736842105263 0.212554
6.10526315789474 0.169333
7.26315789473684 0.133431
8.42105263157895 0.102428
9.57894736842105 0.078808
10.7368421052632 0.059721
11.8947368421053 0.045975
13.0526315789474 0.035754
14.2105263157895 0.02782
15.3684210526316 0.022506
16.5263157894737 0.018184
17.6842105263158 0.015185
18.8421052631579 0.012803
20 0.010953
};
\addlegendentry{\small Learned, $n=6$}
\addplot [thick, green, mark=pentagon*, mark size=2, mark options={solid}]
table {%
-2 0.532392
-0.842105263157895 0.483614
0.315789473684211 0.430659
1.47368421052632 0.37417
2.63157894736842 0.317182
3.78947368421053 0.26362
4.94736842105263 0.213831
6.10526315789474 0.169686
7.26315789473684 0.133071
8.42105263157895 0.102383
9.57894736842105 0.078401
10.7368421052632 0.0598
11.8947368421053 0.045811
13.0526315789474 0.035679
14.2105263157895 0.027875
15.3684210526316 0.022344
16.5263157894737 0.018323
17.6842105263158 0.015001
18.8421052631579 0.012638
20 0.011026
};
\addlegendentry{\small Learned, $n=7$}
\addplot [thick, red, mark=star, mark size=2, mark options={solid,fill opacity=0}]
table {%
-2 0.561143
-0.842105263157895 0.511985
0.315789473684211 0.457838
1.47368421052632 0.395557
2.63157894736842 0.333911
3.78947368421053 0.271931
4.94736842105263 0.214664
6.10526315789474 0.164581
7.26315789473684 0.122581
8.42105263157895 0.088752
9.57894736842105 0.063871
10.7368421052632 0.046171
11.8947368421053 0.034951
13.0526315789474 0.02739
14.2105263157895 0.022469
15.3684210526316 0.019263
16.5263157894737 0.017093
17.6842105263158 0.015183
18.8421052631579 0.014031
20 0.013229
};
\addlegendentry{\small Learned, $n=8$}
\end{axis}

\end{tikzpicture}
        \caption{BLER for \(M=4\) with varying \(n\). The training SNR is \(10\) dB.}
        \label{fig-multi_n_noCsi_M4}
    \end{figure}

\vspace{-0.8cm}
As expected, when \( M=2 \), trading off the code rate results in significantly better performance, especially for \( n=5 \). This scheme is advantageous for systems or applications prioritizing accuracy over information rate. Additionally, the error rate improvements are most notable in low SNR regions (\(<10\) dB). However, with \( n=4 \), a potential issue arises with neural networks. Despite achieving expected training accuracy and loss, the model did not generalize well across a wide range of SNRs during inference. Results for \( M=4 \) is shown in Fig.~\ref{fig-multi_n_noCsi_M4}. Increasing the training SNR led to better results. However, the system's generalization to other SNR values is not as good as in the \( M=2 \) case, with autoencoders performing better than conventional methods only in the region around \( 10 \pm 5 \) dB, where $10$ dB is the training SNR. \\

%\subsubsection{Extension to arbitrary distributions}
% \begin{figure}
%     \centering
%     \input{customDist}
% \caption{PDFs of  distributions considered for learning codebooks in the no-CSI case.}
%     \label{fig:enter-label}
% \end{figure}

\subsubsection{Non-Rayleigh Fading}
\begin{figure}
    \centering
    % This file was created with tikzplotlib v0.10.1.
\begin{tikzpicture}

\definecolor{darkgray176}{RGB}{176,176,176}
\definecolor{deeppink}{RGB}{255,20,147}
\definecolor{lawngreen}{RGB}{124,252,0}
\definecolor{lightgray204}{RGB}{204,204,204}
\definecolor{magenta}{RGB}{255,0,255}
\definecolor{orangered}{RGB}{255,69,0}

\begin{axis}[
legend cell align={left},
legend style={
  fill opacity=0.8,
  draw opacity=1,
  text opacity=1,
  at={(0.03,0.03)},
  anchor=south west,
  draw=lightgray204
},
log basis y={10},
tick align=outside,
tick pos=left,
x grid style={darkgray176},
xlabel={SNR (in dB)},
xmajorgrids,
xmin=-3.1, xmax=21.1,
xtick style={color=black},
y grid style={darkgray176},
ylabel={\(\displaystyle \epsilon_{NoCSI}(2, 2)\)},
ymajorgrids,
ymin=1.21986820836784e-05, ymax=0.645786155091317,
ymode=log,
ytick style={color=black}
]
\addplot [thick, black, mark=*, mark size=1.75, mark options={solid}]
table {%
-2 0.380986
-0.842105263157895 0.354481
0.315789473684211 0.32604
1.47368421052632 0.294986
2.63157894736842 0.261921
3.78947368421053 0.229672
4.94736842105263 0.195391
6.10526315789474 0.165065
7.26315789473684 0.13689
8.42105263157895 0.112092
9.57894736842105 0.091097
10.7368421052632 0.0726
11.8947368421053 0.057513
13.0526315789474 0.045483
14.2105263157895 0.035604
15.3684210526316 0.027865
16.5263157894737 0.021727
17.6842105263158 0.017035
18.8421052631579 0.013508
20 0.010697
};
\addlegendentry{Distribution I, Rayleigh}
\addplot [thick, lawngreen, mark=diamond, mark size=1.75, mark options={solid,fill opacity=0}]
table {%
-2 0.393887
-0.842105263157895 0.37169
0.315789473684211 0.348253
1.47368421052632 0.323923
2.63157894736842 0.296112
3.78947368421053 0.268712
4.94736842105263 0.240751
6.10526315789474 0.213232
7.26315789473684 0.186383
8.42105263157895 0.161081
9.57894736842105 0.138293
10.7368421052632 0.117264
11.8947368421053 0.099059
13.0526315789474 0.082717
14.2105263157895 0.068076
15.3684210526316 0.05663
16.5263157894737 0.0467
17.6842105263158 0.03872
18.8421052631579 0.032455
20 0.026848
};
\addlegendentry{Distribution II, Custom}
\addplot [thick, blue, mark=square*, mark size=1.75, mark options={solid}]
table {%
-2 0.149244
-0.842105263157895 0.123645
0.315789473684211 0.100429
1.47368421052632 0.07946
2.63157894736842 0.061697
3.78947368421053 0.046323
4.94736842105263 0.033899
6.10526315789474 0.024351
7.26315789473684 0.016853
8.42105263157895 0.011691
9.57894736842105 0.00779
10.7368421052632 0.005195
11.8947368421053 0.003327
13.0526315789474 0.002165
14.2105263157895 0.001371
15.3684210526316 0.000844
16.5263157894737 0.000474
17.6842105263158 0.000283
18.8421052631579 0.000184
20 9e-05
};
\addlegendentry{Distribution III, Gamma}
\addplot [thick, deeppink, mark=pentagon*, mark size=2, mark options={solid, fill opacity=0}]
table {%
-2 0.390032
-0.842105263157895 0.364667
0.315789473684211 0.338215
1.47368421052632 0.307881
2.63157894736842 0.276557
3.78947368421053 0.242676
4.94736842105263 0.20977
6.10526315789474 0.177635
7.26315789473684 0.147269
8.42105263157895 0.120936
9.57894736842105 0.09777
10.7368421052632 0.078271
11.8947368421053 0.06233
13.0526315789474 0.049243
14.2105263157895 0.038933
15.3684210526316 0.030604
16.5263157894737 0.024141
17.6842105263158 0.019494
18.8421052631579 0.015636
20 0.012646
};
\addlegendentry{Distribution IV, Gumbel}
\addplot [thick, magenta, mark=o, mark size=1.75, mark options={solid,fill opacity=0}]
table {%
-2 0.091209
-0.842105263157895 0.069646
0.315789473684211 0.051603
1.47368421052632 0.036738
2.63157894736842 0.025606
3.78947368421053 0.017246
4.94736842105263 0.011361
6.10526315789474 0.007522
7.26315789473684 0.004639
8.42105263157895 0.002896
9.57894736842105 0.001806
10.7368421052632 0.0011
11.8947368421053 0.000664
13.0526315789474 0.000427
14.2105263157895 0.000194
15.3684210526316 0.000138
16.5263157894737 9.1e-05
17.6842105263158 4.8e-05
18.8421052631579 3.2e-05
20 2e-05
};
\addlegendentry{Distribution V, Folded Normal}
\end{axis}

\end{tikzpicture}
        \caption{BLER for \(M=2\) and $n=2$ for different distributions of the channel fading coefficients considered in Table~\ref{tab-codebook_distr}. The training SNR is \(10\) dB.}
    \label{fig-diff-distr}
\end{figure}
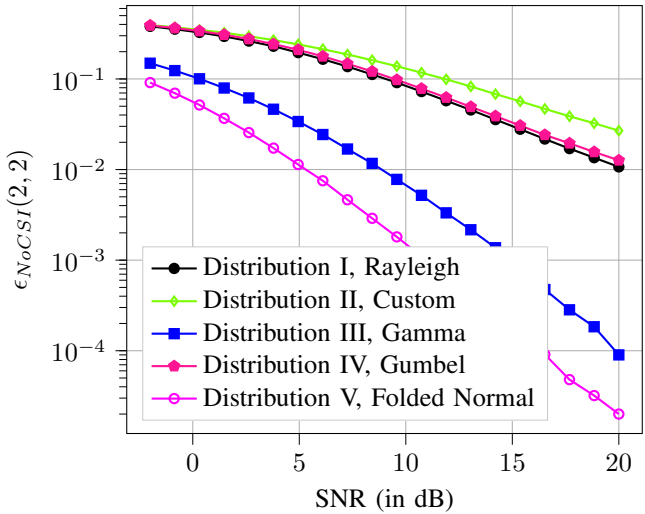

We present the learned codewords and their performance for different distributions on \( h[l] = h_r[l] + j h_i[l] \), in Table~\ref{tab-codebook_distr} and Fig.~\ref{fig-diff-distr}, respectively. \( h[l] \) forms a sequence of i.i.d. random variables.
We consider cases where \( h_r[l] \) and \( h_i[l] \) are i.i.d. with the following probability density functions (PDF):
\begin{itemize}
    \item \textbf{Distribution I (Rayleigh Fading):} Normal, with PDF $f(x ; \mu, \sigma^2) = (\sqrt{2 \pi \sigma^2})^{-1} e^{- (x - \mu)^2 / (2 \sigma^2)}$. The PDF is symmetric around $x=0$. 
    \item \textbf{Distribution II:} Custom, with PDF \( f(x; \lambda) = \lambda e^{-\lambda |x|} / 2 \). The PDF is symmetric around $x=0$. This distribution was obtained by setting $X = Y - Z$, where $Y, Z \overset{i.i.d.}{\sim}\ \rm Exponential(\lambda)$.
    \item \textbf{Distribution III:} Gamma, with PDF \( f(x; k, \theta) = x^{k-1} e^{-x/\theta} (\theta^k \Gamma(k))^{-1} \). The PDF is asymmetric.
    \item \textbf{Distribution IV:} Gumbel, with PDF $f(x; \mu, \beta) = \frac{1}{\beta} e^{ \left( \frac{x - \mu}{\beta} \right) - e^{ \left( \frac{x - \mu}{\beta} \right) }}$. This PDF is asymmetric.
    \item \textbf{Distribution V:} Folded Normal, with PDF $f_X(x; \mu, \sigma) = \frac{1}{\sqrt{2 \pi \sigma^2}} \left( \exp\left(-\frac{(x - \mu)^2}{2 \sigma^2}\right) + \exp\left(-\frac{(x + \mu)^2}{2 \sigma^2}\right) \right), \; x \geq 0$. To obtain samples from this distribution, we sample normal random variable, $Y \sim \mathcal{N}(\mu, \sigma^2)$, and take its absolute value, i.e., $X = |Y|$. This distribution has support  $x \in [0, \infty)$ and is asymmetric.
\end{itemize}

\begin{table}[t]
    \centering
    \caption{Learned codebook for different distributions of channel fading coefficients in the no-CSI case, with \(M=2\) and \(n=2\). We use the following abbreviations: Symmetric (sym.), Asymmetric (asym.), and Support (supp.).}           
    \label{tab-codebook_distr}
    \begin{tabular}{|c|c|p{2.8cm}|}
        \hline
        & Distribution of $h_r$ and $h_l$ & Codewords Learned \\
        \hline
         I. & Rayleigh (sym., supp.: $\mathbb{R}$)  & $[-1.41, 0.0], [0.0, 1.41]$ \\
        \hline
        II. & Custom (sym., supp.: $\mathbb{R}$) &  $[1.41, 0.0], [0.0, 1.41]$ \\
        \hline
        III. & Gamma (asym., supp.: $\mathbb{R}_+$) & $[1.0, -1.0], [-0.99, 1.0]$ \\
        \hline
        IV. & Gumbel (asym., supp.: $\mathbb{R}$) & $[-0.0, -1.4], [-1.4, 0.0]$\\
        \hline
        V. & Folded Normal (asym., supp.: $\mathbb{R}_+$) & $[-1.0, -0.99], [1.0, 0.99]$\\
        \hline
    \end{tabular}
 
\end{table}

 For a fair comparison, the variances of the fading coefficients under each of the considered distributions are normalized to unity by tuning the distribution parameters accordingly. From Table~\ref{tab-codebook_distr}, we observe that \emph{the learned codes are mutually orthogonal when the distribution of the real and imaginary parts of the fading random variable, whether symmetric about zero or not, has support over the entire real line, \(\mathbb{R}\). However, the codes are not mutually orthogonal when the support is the non-negative real line, \(\mathbb{R}_+\).} Further exploration and analysis of this observation will be considered in future work.

%However, from the figure, we observe that performance heavily depends on the specific distribution, despite all of them being chosen with unit variance. 

% {\color{blue}Since the sample size is too small to make a deduction, it is a mere observation that whenever the support of the distribution was $\mathbb{R}_+$, the encodings were not orthogonal to each other.}

%The architecture performs equally well, as demonstrated by both the learned codewords and the performance curves.
% \begin{center}
%     \begin{table}[]
%         \centering
%         \begin{tabular}{|c|c|c|}
%         \hline
%             Distribution & PDF $f(x)$ & Codewords learned \\
%         \hline
%             Normal & 
% $f(x; 0, 0.5) = \frac{1}{\sqrt{\pi}} \exp\left(-x^2\right)$ & $\{[1.4, 0.0014], \\[-0.00073, -1.4]\}$ \\
%         \hline
% %             Gamma  & $f(x; k=3, \theta = \sqrt{0.5/3}) = \frac{x^{k-1} e^{-x/\theta}}{\theta^k \Gamma(k)}$ & $\{[-1.0, 1.0], \\ [1.1, -0.95]\}$\\
% %         \hline
% %         Custom & $f(x;\lambda = 2) = \lambda \frac{e^{-\lambda |x|}}{2} & $\{[0.0016, -1.41], \\ [-1.41, 0.0034]\}$\\
% %         \hline
%         \end{tabular}
%         \caption{Caption}
%         \label{tab:my_label}
%     \end{table}
% \end{center}

\subsection{CSIR-Only Case}
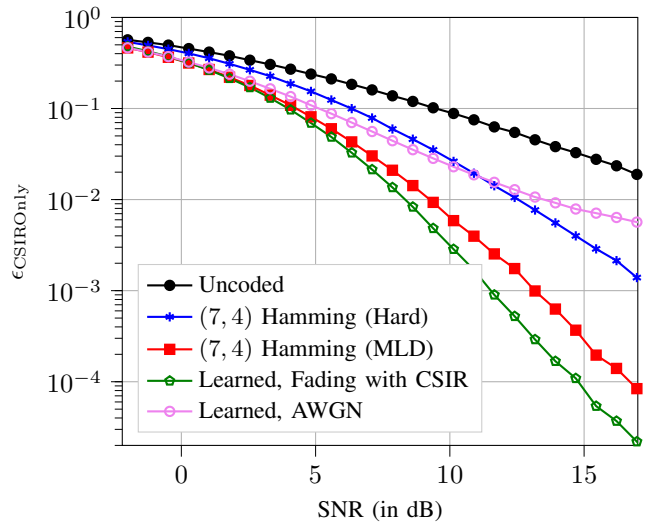
\begin{figure}[t]
    % This file was created with tikzplotlib v0.10.1.
\begin{tikzpicture}

\definecolor{darkgray176}{RGB}{176,176,176}
\definecolor{green}{RGB}{0,128,0}
\definecolor{lightgray204}{RGB}{204,204,204}
\definecolor{violet}{RGB}{238,130,238}

\begin{axis}[
legend cell align={left},
legend style={
  fill opacity=0.8,
  draw opacity=1,
  text opacity=1,
  at={(0.03,0.03)},
  anchor=south west,
  draw=lightgray204
},
log basis y={10},
tick align=outside,
tick pos=left,
x grid style={darkgray176},
xlabel={\small SNR (in dB)},
xmajorgrids,
xmin=-2.2, xmax=17,
xtick style={color=black},
y grid style={darkgray176},
ylabel={\small \(\displaystyle \epsilon_{\mathrm{CSIROnly}}\)},
ymajorgrids,
ymin=.2*1e-4, ymax=0.99886241483954,
ymode=log,
ytick style={color=black}
]
\addplot [thick, black, mark=*, mark size=1.75, mark options={solid}]
table {%
-2 0.56762
-1.24137931034483 0.531888
-0.482758620689655 0.496676
0.275862068965517 0.45474
1.03448275862069 0.416224
1.79310344827586 0.378816
2.55172413793103 0.339156
3.31034482758621 0.304672
4.06896551724138 0.270168
4.82758620689655 0.238868
5.58620689655172 0.211292
6.3448275862069 0.18412
7.10344827586207 0.160024
7.86206896551724 0.137792
8.62068965517241 0.119544
9.37931034482759 0.101796
10.1379310344828 0.088156
10.8965517241379 0.075112
11.6551724137931 0.062784
12.4137931034483 0.054644
13.1724137931034 0.045328
13.9310344827586 0.038132
14.6896551724138 0.032732
15.448275862069 0.027664
16.2068965517241 0.023492
16.9655172413793 0.018896
17.7241379310345 0.016272
18.4827586206897 0.013868
19.2413793103448 0.01192
20 0.009864
};
\addlegendentry{\small Uncoded}
\addplot [thick, blue, mark=asterisk, mark size=1.75, mark options={solid}]
table {%
-2 0.538548
-1.24137931034483 0.492804
-0.482758620689655 0.44804
0.275862068965517 0.402864
1.03448275862069 0.356064
1.79310344827586 0.309352
2.55172413793103 0.26564
3.31034482758621 0.226252
4.06896551724138 0.187052
4.82758620689655 0.154096
5.58620689655172 0.124056
6.3448275862069 0.099688
7.10344827586207 0.07868
7.86206896551724 0.05956
8.62068965517241 0.046044
9.37931034482759 0.035124
10.1379310344828 0.026256
10.8965517241379 0.019456
11.6551724137931 0.014184
12.4137931034483 0.010552
13.1724137931034 0.00768
13.9310344827586 0.005548
14.6896551724138 0.003996
15.448275862069 0.002876
16.2068965517241 0.002136
16.9655172413793 0.001392
17.7241379310345 0.001144
18.4827586206897 0.00078
19.2413793103448 0.000568
20 0.000396
};
\addlegendentry{\small $(7, 4)$ Hamming  (Hard)}
\addplot [thick, red, mark=square*, mark size=1.75, mark options={solid}]
table {%
-2 0.462668
-1.24137931034483 0.41588
-0.482758620689655 0.365024
0.275862068965517 0.31582
1.03448275862069 0.267992
1.79310344827586 0.220332
2.55172413793103 0.17896
3.31034482758621 0.139968
4.06896551724138 0.108344
4.82758620689655 0.081056
5.58620689655172 0.059932
6.3448275862069 0.042976
7.10344827586207 0.030156
7.86206896551724 0.020988
8.62068965517241 0.01424
9.37931034482759 0.009316
10.1379310344828 0.005872
10.8965517241379 0.003964
11.6551724137931 0.002532
12.4137931034483 0.001748
13.1724137931034 0.000992
13.9310344827586 0.000628
14.6896551724138 0.000368
15.448275862069 0.000196
16.2068965517241 0.00014
16.9655172413793 8.4e-05
17.7241379310345 4e-05
18.4827586206897 2.4e-05
19.2413793103448 3.2e-05
20 0
};
\addlegendentry{\small $(7, 4)$ Hamming  (MLD)}
\addplot [thick, green, mark=pentagon, mark size=1.75, mark options={solid,fill opacity=0}]
table {%
-2 0.477838
-1.24137931034483 0.425126
-0.482758620689655 0.371639
0.275862068965517 0.316535
1.03448275862069 0.264166
1.79310344827586 0.21467
2.55172413793103 0.170217
3.31034482758621 0.130283
4.06896551724138 0.096843
4.82758620689655 0.069276
5.58620689655172 0.048683
6.3448275862069 0.032642
7.10344827586207 0.02137
7.86206896551724 0.013594
8.62068965517241 0.008308
9.37931034482759 0.004833
10.1379310344828 0.002852
10.8965517241379 0.001692
11.6551724137931 0.0009
12.4137931034483 0.000523
13.1724137931034 0.000291
13.9310344827586 0.000168
14.6896551724138 0.000109
15.448275862069 5.4e-05
16.2068965517241 3.7e-05
16.9655172413793 2.2e-05
17.7241379310345 1.1e-05
18.4827586206897 1.7e-05
19.2413793103448 7e-06
20 9e-06
};
\addlegendentry{\small Learned, Fading with CSIR}
\addplot [thick, violet, mark=o, mark size=1.75, mark options={solid,fill opacity=0}]
table {%
-2 0.464061
-1.24137931034483 0.416717
-0.482758620689655 0.370912
0.275862068965517 0.323479
1.03448275862069 0.279079
1.79310344827586 0.23791
2.55172413793103 0.199262
3.31034482758621 0.164541
4.06896551724138 0.135119
4.82758620689655 0.10856
5.58620689655172 0.087803
6.3448275862069 0.070083
7.10344827586207 0.055875
7.86206896551724 0.044274
8.62068965517241 0.035291
9.37931034482759 0.028341
10.1379310344828 0.022926
10.8965517241379 0.018697
11.6551724137931 0.015469
12.4137931034483 0.012841
13.1724137931034 0.010655
13.9310344827586 0.009201
14.6896551724138 0.007879
15.448275862069 0.007093
16.2068965517241 0.006359
16.9655172413793 0.00566
17.7241379310345 0.005302
18.4827586206897 0.00474
19.2413793103448 0.004325
20 0.004191
};
\addlegendentry{\small Learned, AWGN}
\end{axis}

\end{tikzpicture}
    \caption{BLER for \(M=4\) with \(n=7\). The training SNR is \(7\) dB. The fading channel considered is a single-input single-output channel.}
    \label{fig:siso_csir}     
\end{figure}
We now consider the CSIR-only case with \( M = 16 \) and \( n = 7 \), as shown in Fig.~\ref{fig:siso_csir}. We train the autoencoder at \(7\) dB SNR for communication over a single-input single-output (SISO) fading channel and compare it with conventional systems: uncoded coherent detection (Uncoded), Hamming \((7, 4)\) hard syndrome decoding with coherent combining (\((7,4)\) Hamming (Hard)), and Hamming \((7, 4)\) soft decoding with maximum-likelihood decoding and coherent combining (\((7,4)\) Hamming (MLD)). From the figure, we observe that the autoencoder-based code performs comparably to the \((7, 4)\) Hamming (MLD), which is the best-performing benchmark, at lower SNRs and outperforms it at higher SNRs.

To demonstrate the optimization of learned codewords, we first trained an autoencoder for an AWGN channel. We then applied these codes to a fading channel by using the encoder to encode the bits, which are passed through the fading channel. At the receiver, we manually perform coherent combining before inputting the signal to the decoder. The results, as shown by the (Learned, AWGN) curve, indicate poor performance in the fading channel.

\balance    
\section{Conclusion and Future Research}
We obtained short-length codewords using a deep learning autoencoder-based approach for both no-CSI and CSIR-only cases. In the no-CSI case, with \( M = 2 \) and \( n = 2 \), the autoencoder learned orthogonal signaling similar to classical techniques under Rayleigh fading. Further exploration with different \( M \) and \( n \), and distributions showed that the learned codes are mutually orthogonal when the distribution of the real and imaginary parts of the fading random variable has support over the entire real line, \(\mathbb{R}\). However, when the support is limited to the non-negative real line, \(\mathbb{R}_+\), the codes are not mutually orthogonal. For the CSIR-only case, codes designed for AWGN channels performed worse in fading channels with optimal coherent detection compared to codes specifically designed for fading channels with CSIR, where the autoencoder jointly learns encoding, coherent combining, and decoding. In both no-CSI and CSIR-only scenarios, these codes perform as well as or better than classical codes of the same block length. Future work includes learning codebooks for MIMO channels with various channel fading distributions and for multi-user communication scenarios.

% Additionally, for the CSIR-only case, deep learning-based codes designed for AWGN channels perform worse when applied to fading channels compared to codes directly designed for fading channels with CSIR. In both no-CSI and CSIR-only scenarios, these codes perform at least as well as classical codes of the same block length.

%Additionally, while the model performs well for AWGN channels, its generalization to fading channels is less effective compared to models trained specifically for AWGN scenarios.

% - With CSIR, the decoder part of the autoencoder learns coherent combining and decoding someway jointly, giving better performance. 

    % References begin here :

    \bibliographystyle{IEEEtran}
    \bibliography{references.bib}
    
\end{document}